\documentclass[mathleft
]{an}
\usepackage{graphicx}
\usepackage{times}
\usepackage{subfigure}
\begin{document}

\Pagespan{789}{}
\Yearpublication{2006}%
\Yearsubmission{2005}%
\Month{11}%
\Volume{999}%
\Issue{88}%

\title{Seismic comparison of the 11 and 2 yr cycle signatures in the Sun}

\author{R.Simoniello \inst{1,2}\fnmsep\thanks{Corresponding author:
\email{rosaria.simoniello@cea.fr}\newline}
\and  K.~Jain\inst{3}
\and S.C.~Tripathy\inst{3}
\and S.~Turck-Chi\'eze\inst{1}
\and W.~Finsterle\inst{2}
\and M.~Roth\inst{3}}
\titlerunning{Quasi-Biennial Periodicity}
\authorrunning{R.Simoniello}
\institute{Laboratoire AIM, CEA-DSM-CNRS-Universit\'e Paris Diderot; CEA, IRFU, SAp, centre de Saclay, F-91191, Gif-sur-Yvette, France
\and 
PMOD/WRC Physikalisch-Meteorologisches Observatorium Davos-World Radiation Center, 7260 Davos Dorf, Switzerland 
\and 
National Solar Observatory, Tucson, AZ 85719, USA
\and Kiepenheuer Institute for Solar Physics, Freiburg}

\received{}
\accepted{}
\publonline{later}


\abstract{The solar magnetic activity consists of two periodic components: the main cycle with a period of 11 yr and a shorter cycle with a period of $\approx$ 2 yrs. The origin of this second periodicity is still not well understood. 
We use almost 15 yrs of long high quality resolved data provided by the Global Oscillation Network Group (GONG) to investigate the solar cycle changes in $p$-mode frequency with spherical degree $\ell=0-120$ and in the range 1600~$\mu$Hz $\le\nu\le$ 3500$~\mu$Hz.
For both periodic components of solar magnetic activity our findings locate the origin of the frequency shift in the subsurface layers with a sudden enhancement in the amplitude of the shift in the last few hundred kilometers. We also show that the size of the shift increases towards equatorial latitudes and from minimum to maximum of solar activity. On the other hand, the signatures of the 2 yr cycle differ from the one of the 11 yr cycle in the magnitude of the shift, as the 2 yr cycle causes a weaker shift in mode frequencies and a slower enhancement in the last few hundreds kilometers.
Based on these findings we speculate that a possible physical mechanism behind the quasi biennial periodicity (QBP) could be the beating between different dynamo modes (dipole and quadrupole mode).} 
\maketitle
\section{Introduction}
Stellar cycles can show all types of periodicities  (Baliunas et al. 1995, Brandenburg, Saar $\&$ Turpin 1998, Ol\'ah $\&$ Strassmeier 2002, Bohm-Vitense 2007) from none to multiple periods.
Even our nearest star shows several periodicities. Sunspot time series dating back to the XVII century reveal that the amplitude and the length of the solar cycle are modulated on different timescales, among which the Gleissberg cycle is the most noticeable one (Koll\'ath $\&$ Ol\'ah 2009). Investigating time series of proxies extending back to $10^{4}$ yr ago, such as the cosmogenic isotopes Be$^{10}$ or C$^{14}$, it is possible to investigate the level of solar activity in the distant past. In combination with sunspot data, these studies revealed intervals of very low solar activity, called grand minima(Usoskin et al. 2007). At the other end of the time scale, there are short and mid-term periodicities, between months and 11-years (Bai 2003). The one attracting a great deal of interest is the quasi-biennial periodicity (QBP) as it appears to modulate mainly all indices of solar activity proxies and is particularly strong over periods coinciding with solar maxima, although it doesn't seem to characterize every solar cycle (Krivova $\&$ Solanki 2002, Vecchio $\&$ Carbone 2009). The discovery of a 1.6 year magnetic activity cycle in the exoplanet host star ι Horologii (Metcalfe  et al. 2010) has been recently reported. This is the shortest activity cycle
so far measured for a solar-type star and it might be related to the mid-timescale magnetic variations recently
identified in HD 49933 and in the Sun from asteroseismic (Garc\'ia et al. 2010) and
helioseismic measurements (Broomhall et al 2009a, Salabert et al. 2010,
  Simoniello et al. 2012). Conversely to solar activity indices, signatures of the QBP
in helioseismic observations have shown its likely persistent nature
throughout solar cycle 23 (Broomhall et al. 2012, Simoniello et al 2012). Several mechanisms have been proposed so far in an
attempt to explain the origin of the QBP, but no clear evidences have been found. We decided to address this issue by investigating the origin and the latitudinal dependence of the shift, because differences or similarities of the seismic properties of the shift over two cycles might help us gaining a deeper insight on the mechanism behind the QBP signal. The findings will also help us improving our understanding on shorter cycles already observed in some stars.
\section{Observations}
\subsection{Determination of mode frequency}
The {\bf G}lobal {\bf O}scillation {\bf N}etwork {\bf G}roup (GONG) consists of six instruments deployed worldwide to provide nearly continuous and stable velocity images of the Sun. The
GONG instrument is based on a Michelson interferometer called a Fourier tachometer. It works by using the Ni line at 676.8~nm, and the observational height is approximately at 213~km. The network started to be fully operating since May 1995. Time series of 36 days (the so-called GONG month) are produced. The mode frequencies for each ($n$~$\ell$~$m$) mode are estimated from three months power spectra using the standard GONG algorithm (Anderson et al. 1990), which fits modes up to $\ell$=150. The peak-fitting algorithm has two types of error flags related to the quality of the fit to a mode (Hill $\&$ Jefferies 1998). The 
individual frequencies of the azimuthal components are afterwards made 
publicly available (http://gong2.nso.edu/archive/). In this work we investigate the temporal evolution of mode frequencies covering a period of observation from May 1995 up to December 2010.
\subsection{Central frequency of the multiplets}
The acoustic modes at the solar surface are described by associated Legendre functions $P^{m}_{l}$(x), where $\ell$ is the degree, $m$ is the azimuthal order, running from -$\ell$ to $\ell$, x is the cosine of the colatitude $\theta$. Rotation and asphericity break the degeneracy among the modes of the same $\ell$ and $m$. The description of the mode is completed by the radial order $n$, with modes of the same $\ell$ and $m$, having different frequencies. The frequencies of modes within an $n$,$\ell$ multiplet are often described by using a polynomial expansion
\begin{equation}
\nu_{n,\ell,m}=\nu_{n,\ell}+\sum a_{j}(n,\ell)P_{j}^{\ell}(m)
\end{equation}
where the basis functions are polynomials related to the Clebsch-Gordan coefficients (Ritzwoller and Lavely 1991) $C_{j0\ell m}^{\ell,m}$ by
\begin{equation}
P_{j}^{\ell}(m)=\frac{\ell\sqrt(2\ell-1)!(2\ell+j+1)!}{(2\ell)!\sqrt(2\ell+1)}C^{\ell,m}_{j0\ell m}
\end{equation}
The term $\nu_{n\ell}$ in this expansion is the so-called central
frequency of the multiplet. 
\subsection{Frequency shift determination}
The temporal variations of the $p$-mode frequencies were defined as the 
difference between the frequencies of the corresponding modes observed 
on different dates and reference values which is the average over the 
years 1995-1996. Since the frequency shifts has a well-known dependency on frequency and mode inertia (Jain, Tripathy $\&$ Bhatnagar 2001), we consider only those modes that are present in all data sets and the shifts are scaled by the mode inertia (Christensen-Dalsgaard $\&$ Berthomieu 1991). The mean frequency shifts is calculated from the following relation
\begin{equation}
\delta\nu(t)=\frac{\sum_{n,\ell,m}{\frac{Q_{n,\ell}}{\sigma^{2}_{n,\ell,m}}}\delta\nu(t)}{\sum_{n,l,m}\frac{Q_{n,\ell}}{\sigma^{2}_{n,l,m}}}
\end{equation}
The weighted averages of these frequency shifts were 
then calculated in two different frequency bands: 
\begin{enumerate}
\item low frequency band 1600~$\mu$Hz$\le\nu\le$ 2500~$\mu$Hz;
\item high-frequency band 2500~$\mu$Hz$\le\nu\le$ 3500~$\mu$Hz.
\end{enumerate}
The frequency dependence analysis might help us localizing the region where the visible manifestation of the 2 yr signal starts to occur. In fact the mode frequency rules the position of the upper turning point of the waves (Chaplin et al. 2001). 
Seismic observations have shown that over the 11 yr activity the size of the shift strongly increases as the mode frequency increases, suggesting that the origin of the shift is predominantly a subsurface phenomenon. In fact, in the Sun's interior the ratio $\beta=\frac{gas\hspace{0.4cm} pressure}{magnetic\hspace{0.4cm} pressure}>>1$ and only close to the surface layers the two terms become of comparable strength. As a consequence we observe a shift in the mode frequencies only very close to the subsurface layers and its amplitude varies along the 11 yr solar magnetic activtiy cycle. It is then important to carry out the same type of investigation over the 2 yr cycle, because if the origin of the shift is due to a mechanism acting at a different depth (such as a second dynamo mechanism close to the subsurface layers), its signature might be revealed by the frequency analysis.
\section{Subsurface analysis and latitudinal dependence of the shift}
\subsection{Localizing the origin of the shift}
We consider the central frequency $\nu_{\ell,n}$ of the ($n$,$\ell$) multiplet, representing a global average of the activity through the frequency shift. We 
investigated the solar cycle changes in $p$-mode central frequency averaged over spherical degree $\ell$=0-120. 
Fig.~\ref{fig:dnu_dependence} compares the solar cycle changes in $p$-mode frequency induced by the two periodic components of solar magnetic activity and the 2-year cycle in the two frequency bands. There are several features underlying the magnitude of the shift over the 2 yr cycles:
\begin{figure}
\begin{center}
\includegraphics[width=3in]{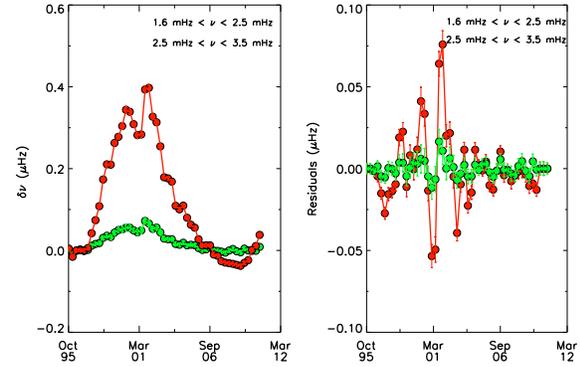}
\caption{The frequency dependence of the shift over solar cycle 23. In the left panel the visible manifestation of the 11 and 2 yr cycles in the low (green) and high (red) frequency band, while in the right panel the visible manifestation of the 2-yr signal once the 11 yr envelope has been subtracted.}\label{fig:11yr_cycle}
\label{fig:dnu_dependence}
\end{center}
\end{figure}
\begin{itemize}

\item it is rather weak;

\item it increases by a factor of $\approx$ 3 from low to high frequency band. This enhancement is smoother compared to the stronger increase of the amplitude of the shift over the 11 yr cycle;

\item it becomes extremely faint over the descending phase of solar cycle 23.
\end{itemize}
This finding is extremely important as it shows that the magnitude of the shift for both periodic components of solar magnetic activity increases with increasing frequency, although the rate of enhancement is reduced over the quasi-biennial cycle. 
\subsection{Selecting latitudes}
The $p$-mode spatial configuration can be described by spherical
harmonics with each mode characterized by its spherical harmonic
degree $\ell$ and azimuthal order $m$. Depending on the ratio
$\frac{m}{\ell}$, the acoustic modes are more sensitive to lower
or higher latitudes. We decided to select three different
latitudinal bands corresponding to equatorial
(0$^{0}$$\le\theta\le$30$^{0}$), mid (30$^{0}$$\le\theta\le$60$^{0}$) and high
latitudes (60$^{0}$$\le\theta\le$90$^{0}$).
In these latitudinal bands the magnitude of the shift over solar cycle 23 is expected to decrease with latitude, as the strong toroidal fields are mainly located in the activity belt between 0$^{0}$$\le\theta\le$45$^{0}$ and the magnitude of the shift over the QBP signal (as shown in Fig.1) is extremely faint. If, instead, a different mechanism is acting behind the 2 yr cycle, we might spot differences in the latitudinal dependence of the shift induced by the 11 and 2 yr periodicity.
\subsection{Latitudinal dependence of the frequency shift}
\begin{figure*}
\begin{center}
\includegraphics[width=5.in]{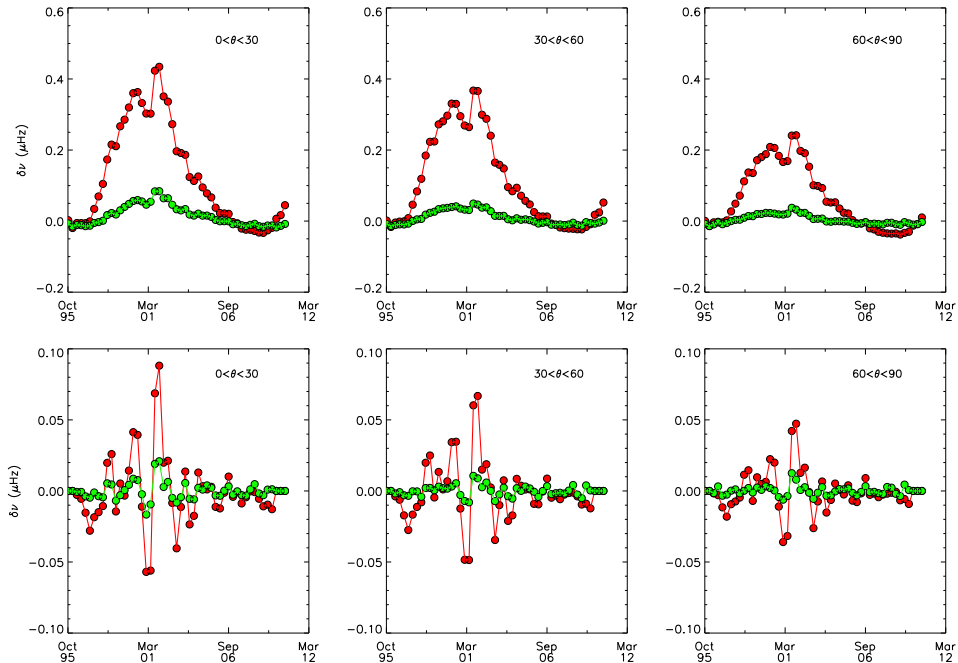}
\caption{The temporal evolution of $p$-mode frequency shifts at different latitudes (from left to right ) over both periodic components of solar magnetic activity cycle (upper panel) and the 2-yr cycle (bottom panel) in low (green) and high (red) frequency bands.}\label{fig:2yr_cycle}
\label{fig:2yr_cycle}
\end{center}
\end{figure*}
Fig.~\ref{fig:2yr_cycle} compares the temporal evolution of the frequency shifts at different latitudes in both frequency bands over the two periodic components of solar magnetic activity (upper panel) and the 2-yr cycle (bottom panel).
 The magnitude of the shift over the QBP shows the following properties:

-  it is modulated by the 11 yr envelope at all latitudes, as it becomes more prominent over periods coinciding with solar maximum;

-  it is more pronounced at equatorial latitudes, where the strong toroidal fields are present;

-  it slightly decreases with increasing latitudes as the shift does over the 11 yr cycle. 

\section{Discussion and Conclusion}
The cyclic behavior of solar and stellar activity is one of the most fascinating questions as the mechanism responsible for it is not yet fully understood. 
 Stellar cycles show all type of periodicities, from none to multi periods. These findings have been explained within the framework of non linear dynamo theories. Also the Sun shows several periodicities, but the one reaching significant level and characterizing all type of solar activity indices is the quasi-biennial periodicity (QBP).
 In the attempt to explain the origin of this second cyclic component, several mechanisms have been proposed so far such as a second dynamo mechanism acting in the subsurface layers (Benevolenskaja 1998a,Benevolenskaja 1998b) or the instability of magnetic Rossby waves acting in the tachocline (Zaqarashvili 2010, Zaqarashvili 2011). Unfortunately the debate is still open as no clear mechanism has been identified and supported by observational evidences. We, then, asked ourselves if the QBP is a phenomenon that might be explained in terms of non linear dynamo theory (as it happens for other stars) or instead we need to invoke a further mechanism.
 To this aim we investigated the properties of the frequency shift over the two periodic components of solar magnetic activity and we found several similarities: i) the observational evidences place the origin of the shift in the subsurface layers, with a sudden enhancement in the size of the shift over the last few hundred kilometers; ii) the magnitude of the shift increases from minimum to maximum of solar activity and towards equatorial latitudes. On the other hand we also found that the signatures of the two cycles differ in the amplitude of the shift as the QBP causes a weaker shift and a slower enhancement in the last few hundred kilometers.
Based on this difference some authors speculated that inside the Sun a second dynamo mechanism is acting (Fletcher et al. 2010, Broomhall et al. 2012), but
if this is indeed true, we wouldn't expect the same latitudinal dependence in the size of the shift over the two cycles (Schou et al. 1998). This might suggest the need to take into account different mechanisms to explain the origin of the QBP (Simoniello et al. 2012). To this aim it will be important to complete the investigation with a depth dependence analysis of the shift in the subsurface layers. If this further analysis will not show any difference over the two cycles, then we should start exploring the possibility that the QBP signal might be the result
of the beating between different dynamo modes, such as dipole and quadrupole modes. Within this thory, the secondary cycle is shown to have lower amplitude as we observe in the Sun (Brandenburg 1989, Fluri and Beryugina 2004, Sokoloff 2005, Tobias et al. 2005, Moss 2008). Furthermore, there is evidence that a quadrupole-like field was important during the Maunder Minimum and there is also a contemporary evidence that there is still a significant quadrupolar component (Pulkkinen et al. 1999). 
\acknowledgements
This work utilizes GONG data obtained by the NSO
Integrated Synoptic Program (NISP), managed by the
National Solar Observatory, which is operated by AURA,
Inc.  under a cooperative agreement with the National
Science Foundation.  The data were acquired by instruments
operated by the Big Bear Solar Observatory, High Altitude
Observatory, Learmonth Solar Observatory, Udaipur Solar
Observatory, Instituto de Astrofisica de Canarias, and
Cerro Tololo Interamerican Observatory."

\end{document}